\documentclass{article}

\usepackage[preprint,nonatbib]{neurips_2020}

\usepackage[utf8]{inputenc} 
\usepackage[T1]{fontenc}    
\usepackage{hyperref}       
\usepackage{url}            
\usepackage{booktabs}       
\usepackage{amsfonts}       
\usepackage{amsmath}
\usepackage{nicefrac}       
\usepackage{microtype}      
\usepackage{multirow}
\usepackage{array}
\usepackage{bm}
\usepackage{xspace}
\usepackage[utf8]{inputenc} 
\usepackage[T1]{fontenc}    
\usepackage{hyperref}       
\usepackage{url}            
\usepackage{booktabs, arydshln}       
\usepackage{tikz}
\usepackage[ruled,linesnumbered]{algorithm2e}
\usepackage{subcaption}
\usepackage{enumerate}
\usepackage{enumitem,amsmath,amssymb,bbm}
\usepackage{titlesec}
\titlespacing\section{0pt}{0pt plus 0pt minus 0pt}{0pt plus 0pt minus 0pt}
\titlespacing\subsection{0pt}{0pt plus 0pt minus 0pt}{0pt plus 0pt minus 0pt}
\titlespacing\subsubsection{0pt}{0pt plus 0pt minus 0pt}{0pt plus 0pt minus 0pt}

\definecolor{cell}{HTML}{999999}
\definecolor{lightgray}{HTML}{F0F0F0}  
\definecolor{rowbackground}{HTML}{F9F9F9}

\definecolor{rest-color}{HTML}{457CFC}
\definecolor{expert-color}{HTML}{319B51}
\definecolor{baseline-color}{HTML}{BDBDBD}

\definecolor{rest-color}{HTML}{457CFC}
\definecolor{expert-color}{HTML}{319B51}
\definecolor{baseline-color}{HTML}{BDBDBD}

\newcommand{\mname}{\texttt{EMIXER}\xspace}

\title{\mname: End-to-end Multimodal X-ray Generation via Self-supervision}

\author{%
  Siddharth Biswal \\
  Georgia Tech \\
  \texttt{sbiswal7@gatech.edu} \\
  \And
  Peiye Zhuang \\
  UIUC \\
  \texttt{peiye@illinois.edu} \\
  \AND
  Ayis Pyrros\\
  Dupage Medical Group \\
  \texttt{ayis@ayis.org} \\
  \And
  Nasir Siddiqui \\
 Dupage Medical Group \\
   \texttt{nsiddiqui@gmail.com} \\
  \And
  Sanmi Koyejo \\
  UIUC \\
   \texttt{sanmi@illinois.org} \\
  \And
  Jimeng Sun \\
  UIUC \\
   \texttt{jimeng@illinois.edu} \\
}

\begin{document}

\maketitle

\begin{abstract}
Deep generative models have enabled the automated synthesis of high-quality data for diverse applications. However, the most effective generative models are specialized to data from a single domain (e.g., images or text). Real-world applications such as healthcare require multi-modal data from multiple domains (e.g., both images and corresponding text), which are difficult to acquire due to limited availability and privacy concerns and are much harder to synthesize. 
To tackle this joint synthesis challenge, we propose an End-to-end MultImodal X-ray genERative model (\mname) for jointly synthesizing x-ray images and corresponding free-text reports, all conditional on diagnosis labels. \mname is an conditional generative adversarial model by 1) generating an image based on a label, 2) encoding the image to a hidden embedding, 3) producing the corresponding text via a hierarchical decoder from the image embedding, and 4) a joint discriminator for assessing both the image and the corresponding text. 
\mname also enables self-supervision to leverage vast amount of unlabeled data. Extensive experiments with real X-ray reports data illustrate how data augmentation using synthesized multimodal samples can improve the performance of a variety of supervised tasks including COVID-19 X-ray classification with very limited samples. The quality of generated images and reports are also confirmed by radiologists.
We quantitatively show that \mname generated synthetic datasets can augment X-ray image classification, report generation models to achieve $5.94\%$ and $6.9\%$ improvement on models trained only on real data samples. Taken together, our results highlight the promise of state of generative models to advance clinical machine learning. 
\end{abstract}

\section{Introduction}

While clinical applications of supervised machine learning algorithms continue to advance, their impact is stifled by the limited amount of available labeled clinical data. This issue only made more dire by applications such as radiology report generation for medical images, which require paired data jointly across images, clinical notes, and diagnosis labels. Data sharing across healthcare organizations and institutions remains difficult, often due to legal and privacy concerns~\cite{mcguire2008confidentiality, filkins2016privacy}. 
On the other hand, generative modeling has improved dramatically in the past few years. While early Generative Adversarial Networks (GANs) could only synthesize low-resolution grayscale images~\cite{goodfellow2014generative}, state-of-art generative models can now synthesize diverse high-quality and high-resolution images ~\cite{brock2018large, karras2017progressive, karras2019style, karras2019analyzing}. GANs and related generative models have been applied to various domains such as computer vision
\cite{brock2018large, karras2017progressive}, natural language processing \cite{dai2017good,fedus2018maskgan}, time-series synthesis \cite{brophy2019quick}, semantic segmentation \cite{dong2017semantic,luc2016semantic}, among others. This manuscript explores using generative models to address the challenge of limited data in machine learning for clinical applications. 
We explore a variety of applications, with a focus on using synthetic data to augment real datasets -- increasing the amount of the data and labels available~\cite{choi2017generating}, thereby improving downstream model performance. 


We focus on X-rays as are a primary diagnostic tool in many clinical workflows, most importantly in radiology, and are used for detecting pneumonia, bone fracture, and cancer\cite{rajpurkar2017chexnet, gulshan2016development}. Recent research efforts have shown promise for lung cancer detection in radiology, prostate cancer in pathology, and differential diagnoses in dermatology ~\cite{ardila2019end,fujisawa2019deep,arvaniti2018automated,mohamed2018deep}. Most recently, X-rays have been employed for the coronavirus diagnosis and prognosis~\cite{jacobi2020portable}. Along with X-rays, associated reports written by clinicians are the primary communication between patients and doctors~\cite{schwartz2011improving, kahn2009toward}. Several deep learning based X-ray image to report writing method been proposed ~\cite{jing2017automatic,jing2020show,li2018hybrid}. Researchers have proposed generative models for clinical data~\cite{choi2017generating}. However, existing methods are limited to a single modality -- images, or clinical reports only.
Thus, current generative models are not able to produce high quality multimodal synthetic datasets, which is the focus of this paper. 
This manuscript investigates an end-to end approach for generating multimodal X-ray images and text reports which are essential for the radiology applications. To this end, our work addresses the following challenges.  
\begin{itemize}[leftmargin=*]
    \item \textbf{Multimodal generation of images and corresponding reports}: Multimodal generative models are difficult to train compared to single-mode modal generative models \cite{liu2016coupled, isola2017image, zhu2017toward, zhu2017unpaired, choi2018stargan, choi2019stargan}. In the past few years, there have been multiple attempts at developing models that can generate multiple modalities at the same time \cite{pu2018jointgan}. In particular, text synthesis using generative models has proven to be extremely challenging -- most likely because because discrete text tokens are not differentiable --  making it more difficult to train GANs. We show that using an end-to-end approach, combines with appropriate text embeddings can overcome these issues. 
    
    \item \textbf{Generative model training with limited labels}:
    Generative models typically require large quantities of high-quality labeled data for training. However, labels are scarcely available in real-world applications such as medical domain. This renders training of high-quality generative models challenging. We present successful results with limited labeled X-ray data along with large amount of unlabeled X-ray data, and conjecture about properties of X-rays which make this feasible. 
    
    \item \textbf{Difficultly of data augmentation with limited  data:}
    The task of training a generative model for classifier augmentation~\cite{huang2018auggan,antoniou2017data} is particularly challenging in the case of rare diseases or new phenotypes, as the limited amount of labels renders training of generative models difficult. For example, in the case of the COVID-19 pandemic, the amount of available X-ray data and labels are extremely low. Given the limited labels, training high-quality generative models to augment the original dataset is a challenge. Pretraining models of large and diverse augmented data can potentially provide robust embeddings for new phenotypes.
\end{itemize}

We propose \mname, an end-to-end multimodal generative model that can generate paired chest X-ray images and corresponding reports simultaneously, conditioned on diagnosis labels. Our primary contributions are summarized in the following.

\begin{itemize}[leftmargin=*]
\item \textbf{Multimodal X-ray image and report generation.}
We show that \mname generates high-quality X-ray images and corresponding reports.
Multiple radiologists scored average 7.340/10 for synthetic data and 7.825/10 for real data on their realisticness and quality. Furthermore, \mname generated synthetic datasets used to augment X-ray image classification models lead to up to  $5.94\%$ improvement in classification accuracy compared to models trained on real X-ray images only. Similarly, \mname augmented paired X-ray image and report datasets improve X-ray report generation models up to $6.9\%$ as measured by the CIDEr scores.

\item \textbf{Learning high-quality generative models from limited samples.}
\mname uses self-supervision to enable learning of high-quality generative models from limited labels. We show that even with $30\%$ of the original labels, \mname can outperform baselines with the 100\% labeled data in terms of image classification and report generation tasks.
 
\item \textbf{Improved classification of COVID-19 chest X-rays via data augmentation.}
We utilize the pre-trained model of \mname with augmenting classification models, applied to the automated diagnosis of COVID-19 from X-ray images. Our results show $11\%$ improvement in predictive accuracy than the one without using pre-trained \mname model.
\end{itemize}

\section{Related Work}
\noindent \textbf{Generative models.} 
In the past few years, there has been great progress in the area of generative modeling of complex imaging data. Since the introduction of the Generative Adversarial Networks (GAN), there have been many variants proposed such as DCGAN, Progressive GAN, Self-supervised GANs \cite{goodfellow2014generative, karras2017progressive, radford2015unsupervised, dai2017towards}, among others. In addition to GANs, other types of generative models are also quite widely used such as Flow Models, Autoregressive Models, and variational autoencoders \cite{kingma2013auto,kingma2018glow, dinh2014nice, dinh2016density}. Flow Models uses a stack of invertible transformations to a sample from prior distributions, thus can compute the exact log-likelihoods of observations. Autoregressive models factorize the distribution over observations into a sequence of conditional distributions (e.g. over pixels for images), then process each component in sequence ~\cite{oord2016pixel,van2016conditional}. For image generation applications, GAN-based models produce among the photo-realistic images. However, the training of GAN models can be quite challenging with known issues such as mode collapse and instability in convergence \cite{salimans2016improved}. There have been many works to improve upon these challenges, e.g., by changing the objective function~\cite{arjovsky2017wasserstein}.  Some other research efforts have focused on constraining the discriminator through gradient penalties or normalization \cite{miyato2018cgans}. BigGAN~\cite{zhang2018self, brock2018large} adds the self-attention block, and ProGAN  considers training a single model across a sequence of increasing resolutions~\cite{karras2017progressive}. While there is a lot of effort in modeling single modalities especially images, there is a dearth of research on multimodal image and text generation.This work addresses the challenge of multimodal joint generation of image and text.

\noindent \textbf{Medical report generation.} 
Deep learning based image classification has been successfully applied to many different types of medical image classification tasks such as diabetic retinopathy classification, X-ray classification, cancer detection from cell images, and X-ray based bone classification~\cite{wang2018tienet, gulshan2016development, milletari2016v}, among other applications. Similarly, different image segmentation algorithms have been very successfully applied to medical images to identify different organs and diseases. There has been progress in the task of automated report generation for medical images such as X-rays \cite{liu2019clinically}.  Existing applications of machine learning to clinical tasks must address a variety of challenges such availability of large datasets. 


\section{Methods}
\subsection{Problem Definition}
We begin by introducing notations. We denote real chest  X-ray images as $\mathbf{I_{n}} \in \mathbb{R}^{l \times l}$ where $l\times l$ is the size of the image, text X-ray reports as $\mathbf{S_{n}}$ and labels as $\mathbf{y_{n}} \in \{0,1\}^{k}$ for $n$th data sample.
The X-ray report $\mathbf{S_{n}}$ contains a sequence of sentences $\mathbf{S_{n}}$ = $[\bm{s_{1}^{n}} \ldots \bm{s_{T}^{n}}]$, where the report length $T$ may vary. Sentence $\bm{s_{t}^{n}}$ consists of sequence of words $\bm{s_{t}^{n}}$ = $[\bm{a_{t,1}^{n}}, \bm{a_{t,2}^{n}}, \ldots ]$ where $\bm{a_{t,j}^{n}}$ $j$-th word represented as one-hot vectors in the  sentence $t$ of document $n$.
The dataset, denoted as $\mathcal{E}$ is a combination of images $\mathbf{I_{n}}$, reports $\mathbf{S_{n}}$ and labels $\mathbf{y_{n}}$ denoted as $\mathcal{E}=\left\{\mathbf{I_{n}}, \mathbf{S_{n}}, \mathbf{y_n} \right\}_{n=1}^{N}$.
\mname generates synthetic dataset that consists of synthetic X-ray images $\mathbf{\hat{I}_{n}}$, synthetic report $\mathbf{{\hat{S}_{n}}}$ conditioned on class labels. 
We train an end-to-end generative model which consists of an X-ray image generator $\operatorname{G}$, X-ray image discriminator $\operatorname{D_{\text{image}}}$, X-ray report discriminator $\operatorname{D_{report}}$, and an X-ray image to report decoder $\operatorname{F}$. Each of these components is a neural network that are trained jointly to produce paired X-ray images and clinical reports conditioned on diagnosis labels.


\begin{figure*}[!t]
\centering
    \includegraphics[width=1\textwidth]{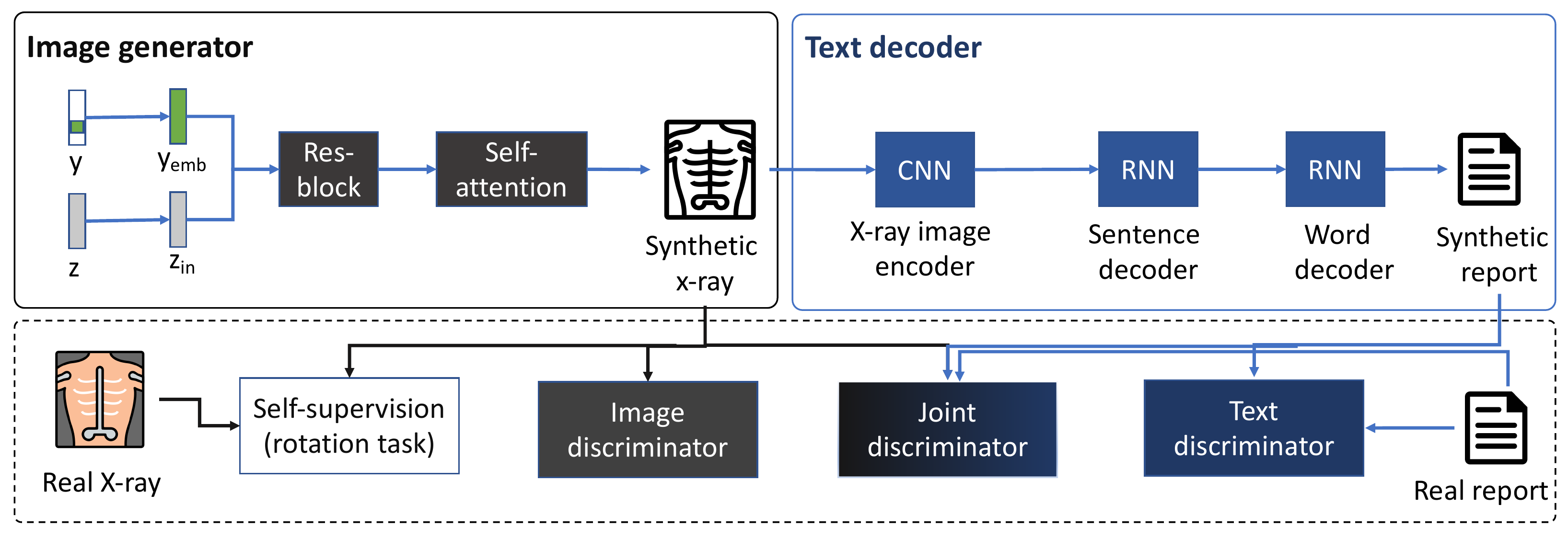}
    \caption{An overview of \mname generator framework}
    \label{fig:framework_fig}
\end{figure*}




\subsection{The \mname Model}
We describe primary components of \mname in this section. As illustrated in Fig.~\ref{fig:framework_fig}, \mname is composed of four different trainable networks: (a) \textbf{Image generator}: This image generator synthesizes X-ray images from a prior noise distribution conditioned on label information (b) \textbf{Image to report decoder}: An image to report decoder produces a text report from X-ray image (c) \textbf{Image Discriminator}: This discriminator is tasked with discriminating between real and synthetic X-ray images (d) \textbf{Text Discriminator}: This text discriminator distinguishes between real and synthetic X-ray reports (e) \textbf{Joint discriminator}: The joint discriminator combines the embedding of X-ray images and text to discriminate between real and synthetic embeddings.

\subsubsection{X-ray Image Generator ($\operatorname{G}$)}
An X-ray image generator is a deep neural network that accepts two inputs; a noise vector $\boldsymbol{z} \in \mathbb{R}^{d_{z}}$ and class information $\mathbf{y}$ represented as one-hot vector. 
First, we split the noise vector $\mathbf{z}$ to obtain $\mathbf{z_{spl}} \in \mathbb{R}^{20}$ vectors.
The vectors $\mathbf{z_{spl}}$ is passed through a linear layer to obtain $\mathbf{z_{in}}$,  \(\mathbf{z_{in}} = \mathbf{W_l}\mathbf{z_{spl}}+b_{l} \). We embed the class information $\mathbf{y}$ via a linear layer to obtain $\mathbf{y_{emb}}  \in \mathbb{R}^{128}$. $\mathbf{z_{in}}$ concatenated with  $\mathbf{y_{emb}}$ is passed through three layers of $\operatorname{residual-block}$ which applies batch-normalization with deconvolution operation, \(\mathbf{res_{out}} = \operatorname{res-block}(\mathbf{z_{in}}, \mathbf{y_{emb}})\)~\cite{he2016deep}. The output $\mathbf{res_{out}}$ is passed through a self-attention block which applies applies a $1 \times 1$ convolution operation with softmax to obtain intermediate feature vectors which are combined with the original input to compute the $\operatorname{att-block}$, \(  \mathbf{\mathbf{att_{out}}}=\operatorname{self-att-block}(\mathbf{res_{out}})\). Finally this output $\operatorname{self-att-block}$ is passed through another $\operatorname{res-block}$ to obtain the $\hat{\mathbf{I}}$ as the output of image generator. Taken together, the generator network can be abstracted as the following \(\hat{\mathbf{I}} = G\left(\mathbf{z}, \mathbf{y}\right)\).  We provide implementation details of $\operatorname{res-block}$, $\operatorname{self-att-block}$ blocks in the supplement. 





    

\subsubsection{X-ray Report Generator ($\operatorname{F}$)}

The image is fed through an image encoder convolutional neural network($\operatorname{CNN}$) to obtain a feature representation. These feature vectors are passed to a sentence decoder $\operatorname{RNN}$ to recurrently generate topic vectors for each sentence. These topic vectors are used by a word decoder to generate the words for each sentence as \(  \mathbf{\hat{S}} = \operatorname{F}(\mathbf{\hat{I}})\).  


\textbf{X-ray image encoder} 
 Specifically, given an image $\mathbf{I}$, we first extract its features $\overline{\mathbf{v}} \in \mathbb{R}^{512}$ from an intermediate layer of a $\operatorname{CNN}$, $\overline{\mathbf{v}} =\operatorname{CNN}(\mathbf{I}) $. We use a pretrained DenseNet-121 as the $\operatorname{CNN}$ model trained on a different chest X-ray dataset \cite{huang2017densely}. Note that this $\operatorname{CNN}$ is different from the  $\operatorname{CNN}$ used in the image discriminator $\operatorname{D_{image}}$. The report generator module is composed of a sentence decoder and word decoder RNN which are described below.

 \textbf{Sentence decoder RNN}: Given the X-ray image features $\overline{\mathbf{v}}$ extracted by the $\operatorname{CNN}$, a sentence decoder is used generate topic vectors $\mathbf{t_i}$. We employ a Long-Short Term Memory network (LSTM) to compute the hidden state as  $\mathbf{h}_{i} =\operatorname{LSTM(\overline{\mathbf{v}};\mathbf{h}_{i-1})} $. We use the hidden states in two ways: First, we project the hidden state $\mathbf{h_i}$ through a linear layer and logistic layer to get probability distribution $\mathbf{u_i}$ over two states {CONTINUE = 0, STOP = 1}. Second, we also feed $\mathbf{h_i}$ through three-layered fully connected network to get a topic vector $\mathbf{t_i}$for $i$th sentence in the report, $ \mathbf{t_i} = \mathbf{W_\text{to}}\mathbf{h_{i}}+b_{\text{to}} $. 

 \textbf{Word decoder RNN}: The words for each individual sentence are generated by a word decoder which is a trainable three-layer LSTM. The sentence topics $\mathbf{t_i}$ generated by the sentence decoder are combined with the $\text{<START>}$ token as input for the first and second input to the word LSTM. In subsequent steps, we provide the hidden state of the last LSTM layer to predict a distribution over the words in the vocabulary. The hidden state $\mathbf{h}_{\text {word}} \in \mathbb{R}^{H}$ of the word LSTM is directly used to predict the distribution over words:
$p\left(\text{word} | \mathbf{h}_{\text{word}}\right) = \mathrm{softmax} \left(\mathbf{W}_{\text{out}} \mathbf{h}_{\text{word}}\right)$
where $\mathbf{W}_{\text {out}}$ is the parameter matrix. Finally, after the word decoder generates the word sequences, we concatenate all the generated sequence to obtain the final report.

\subsubsection{Discriminator ($\operatorname{D}$)} \mname uses three discriminators, an image discriminator, a report discriminator and joint embedding discriminator to ensure image and report consistency of the synthetic data. The image discriminator measures whether the generated image $\mathbf{\hat{I}}$ matches the image distribution of real X-ray images, and the report discriminator $\mathbf{\hat{S}}$ discriminates between the real and synthetic X-ray reports.

\textbf{X-ray Image discriminator ($\operatorname{D_{image}}$)}: We use a convolutional neural network discriminator for X-ray images which are fed real and synthetic X-ray images for classification. 
The discriminators use a ResNet architecture where the input image is passed through multiple layers of ResBlocks, where ResBlocks are composed of $3\times3$ convolution with $\operatorname{ReLU}$ layers~\cite{he2016deep}. This image discriminator can be represented as $D(\mathbf{I},y)=c_{\mathrm{rf}}(\tilde{D}(\mathbf{I}))+P(\tilde{D}(\mathbf{I}), y)$ where $P(\tilde{v}, y)=\tilde{x}^{\top} W y$ is a linear layer with weight matrix $W$ applied to to image feature $v$ and one-hot encoded label $y$. $c_{\mathrm{rf}}$ is a linear classifier tasked with detecting if the provided sample is real or fake.

\textbf{X-ray Report Discriminator ($\operatorname{D_{report}}$)}: We use a X-ray report discriminator which  classifies a given X-ray report as real or fake. X-ray reports generated from the decoder and real X-ray reports are passed as input  discriminator. We employ a LSTM to to extract text embeddings from given X-ray report $\mathbf{S}$, \(\mathbf{p} = \operatorname{LSTM} (\mathbf{S}) \)  \cite{cho2014learning}. These report embeddings $\mathbf{p}$ are passed through multi-layer linear layers with softmax layer to obtain $\mathbf{y_{\mathrm{r/f}}}$. The report discriminator can be abstracted as to discriminate between real or fake report embedding as \( \mathbf{y_{\mathrm{r/f}}} = {\operatorname{D_{report}}(\mathbf{\hat{e}})} \). We provide further details of the implementation in the supplementary section.



\textbf{Joint Discriminator for X-ray images and Reports ($\operatorname{D_{joint}}$)}: 
Along with the image discriminator and report discriminator, we also use a joint embedding discriminator. We hypothesize that as the X-ray images and reports are dependent upon each other, a joint multimodal embedding discriminator provides further guidance to the generator network for generating higher quality images and reports. This joint embedding discriminator is designed to discriminate real joint embeddings from fake joint embeddings. The joint embedding discriminator first obtains image features  $\mathbf{I_{emb}}$ from the X-ray images using a $\operatorname{CNN}$ before the pooling layer. The text-reports are provided as input to an LSTM. The last hidden vector of the LSTM is passed through a linear layer to obtain report embedding $\mathbf{S_{emb}}$. The image feature vector $\mathbf{I_{emb}}$  and report embedding  $\mathbf{S_{emb}}$ are concatenated together to form the joint embedding $\mathbf{C_{joint}}$. This joint embedding is passed through linear layers to obtain probability of real or fake embedding. This discriminator can be abstracted as  \( \mathbf{y_{\mathrm{r/f}}} = {\operatorname{D_{joint}}(\mathbf{\hat{C}_{joint}})}\). 


\textbf{Learning}:
Previous works have shown that self-supervision guide the classifier to learn useful data representation by detecting auxiliary information such as rotation angles. When applied to image classification, typically images are rotated and the angle of rotation is provided as the artificial label. In this rotation task, the self-supervised task is to predict the angle of rotation of an image. We use $\mathcal{R} = \{0,90, 180, 270\}$ rotation angles. Image $\mathbf{I}$ is rotated by $r$ degrees is denoted as $\mathbf{I}^{r}$ and $Q_{D_{\text{image}}}\left(R=r | \mathbf{I}^{r}\right)$ is probability distribution over the rotation angles. The \mname framework corresponds to a constrained minimax game given by where the value function $V$ is given by
\vspace{-0.2cm}
\[
\begin{aligned}
V\left(G, D_{*}\right) &=
\mathbb{E}_{\mathbf{x}_{1} \sim p_{\mathbf{X}_{1}}}\left[-\log D_{\text{image}}\left(\mathbf{\hat{I}}\right)\right]
+ \mathbb{E}_{\mathbf{z} \sim p_{\mathbf{Z}}}\left[-\log \left(1-D_{\text{image}}\left(G(\mathbf{z})\right)\right)\right] \\
& + \mathbb{E}_{\mathbf{x}_{1} \sim p_{\mathbf{X}_{1}}}\left[-\log D_{\text{report}}\left(\mathbf{\hat{S}}\right)\right] + \mathbb{E}_{\mathbf{z} \sim p_{\mathbf{Z}}}\left[-\log \left(1-D_{\text{report}} (F \left(G(\mathbf{z})\right)\right)\right]  \\ 
& + \mathbb{E}_{\mathbf{x}_{1} \sim p_{\mathbf{X}_{1}}}\left[-\log D_{\text{joint}}\left(\mathbf{\hat{I}}, \mathbf{\hat{S}}\right)\right]  \\ 
& + \mathbb{E}_{\mathbf{z} \sim p_{\mathbf{Z}}}\left[-\log \left(1-D_{\text{joint}} (F \left(G(\mathbf{z})\right), G(\mathbf{z})\right)\right]  \\ 
& + \alpha \mathbb{E}_{\boldsymbol{x} \sim P_{G}} \mathbb{E}_{r \sim \mathcal{R}}\left[\log Q_{D_{\text{image}}}\left(R=r | \mathbf{I}^{r}\right)\right]
\end{aligned} 
\]
where $G$, $D_{\text{image}}$, $D_{\text{report}}$, $D_{\text{joint}}$, $F$ are the image generator, image discriminator, report discriminator and image to report decoder, respectively. \mname can be trained by back propagation with the alternating gradient update steps. The details of the learning algorithm are given in the supplementary materials.

\section{Experiments}
In this section, we perform extensive evaluations to measure the effectiveness of \mname for paired chest X-ray image and report generation. We empirically show that (1) our proposed model can generate high-quality X-ray images and reports (2) \mname with self-supervised loss can match the generated sample quality of the conditional models using only fraction of labels (3) \mname can be used to augment datasets in limited label settings such as COVID-19 chest X-ray detection. 


\subsection{Datasets}
We perform experiments on MIMIC-CXR dataset, one of the largest X-ray datasets containing 377,110 X-ray images and corresponding reports~\cite{johnson2019mimic}. MIMIC dataset contains 377,110 chest X-rays associated with 227,827 imaging studies sourced from the Beth Israel Deaconess Medical Center between 2011-2016. The labels extracted from the reports contain 14 different unique classes. We resize the images to $128 \times 128\times 3$ as done in previous work~\cite{miyato2018cgans}. 



\subsection{Evaluation Metrics}
We perform quantitative and qualitative experiments: (a) For classification experiments, we used accuracy and AUC as classification metrics. 
We use CIDEr, BLEU scores for image captioning experiments~\cite{vedantam2015cider,papineni2002bleu}. 
(b) To evaluate X-ray image quality, we use the Fréchet Inception Distance (FID) scores. We use a special pre-trained Inception network on chest X-ray images. We have provided further details in the supplement. (c) We qualitatively evaluate the generated X-ray images and reports. For this, we present randomized pairs of real or synthetic X-ray images and reports to clinical experts for evaluation (they do not know if the presented sample is real or synthetic). The clinical experts were asked to provide a numerical quality score between 1-10 (10 being the best) for each sample.



\subsection{Models}

\noindent \textbf{JointGAN}: JointGAN trains multiple generators and a single softmax-based critic, all jointly trained via adversarial learning\cite{pu2018jointgan} to generate joint data distributions. 
\noindent \textbf{CoGAN}: CoGAN learns separate generators for two different domains with tied weights on the first few layers for shared latent representations~\cite{liu2016coupled}. 
\noindent \textbf{Single Modal Image GAN with text decoder(SM-GAN)} In this setup, we use a GAN model to generate X-ray images. These X-ray images are passed to a text decoder which produces text reports corresponding to the synthetic chest X-rays.
\noindent \textbf{\mname}:  We compare these baselines against our method which is a self-supervised generative model with multiple discriminators for each modality. The final loss for our discriminator is the combination of adversarial loss of both the generators and joint embedding. 

\subsection{Experimental Results and Discussion}
Our experiments aim to answer the following questions.
\begin{itemize}[noitemsep]
    \item Can \mname generate high quality X-ray images?
    \item Can \mname generate high quality pairs of X-ray images and reports? 
    \item Can \mname learn a high quality generative model from limited samples?
    \item Can \mname be used to improve COVID X-ray classification?
\end{itemize}
\subsubsection{Image quality evaluation: Is \mname capable of generating high quality X-ray images?}
One of the primary applications of generative models is data augmentation to increase sample size and improve downstream model performance. We use the baselines and \mname to augment the real X-ray images and evaluate the improved quality of the datasets by using these augmented datasets for X-ray image classification. 

\textbf{X-ray image classification setup}: We trained two separate X-ray image classification models on real X-ray images and synthetic X-ray images. We hypothesize that good generative models should generate images that resemble real data and can be used to train a classification model. These classification models are evaluated on held out real X-ray images. This setup evaluates the performance of the classification model for five different classes of diseases related to the X-ray images.  In this experiment, we report accuracy and AUC for classification scores in Table~\ref{tab:data_augmentation_exp}, where we increase the dataset size by augmenting the real data with generated X-ray images. We use 100k real X-ray images and gradually increase the augmented dataset size by adding synthetic X-ray images up to 600k. We notice improved performance of these image classification model by up to $5.94\%$ compared to real X-ray images, and $3.6\%$ improvement compared to the best baseline. This highlights that \mname is able to generate synthetic X-ray images which are able to augment the real dataset to improve the classification performance.

\begin{table}[ht]
\centering
  \caption{Comparison of X-ray report generation model performance with real and augmented dataset; In this table R indicates real data samples, S indicates Synthetic data samples} 
  \label{tab:data_augmentation_exp}
  \resizebox{1\columnwidth}{!}{
  \begin{tabular}{clccccccc}
    \toprule
      &  &   \multicolumn{2}{c}{Image Classification} & \multicolumn{5}{c}{Report Generation} \\ 
    \midrule
    Dataset & Method & AUC & ACC & CIDEr & BLEU-1 & BLEU-2 & BLEU-3 & BLEU-4 \\
    \midrule
    \multirow{1}{*}{Only real data}
    &\multirow{1}{*}{R 100k } &   $.824 \pm 0.0034 $ & $.846 \pm 0.0041 $ &  $.712 \pm 0.0014 $ & $.253 \pm 0.0024 $ & $.198 \pm 0.0034 $ & $.095 \pm 0.0041 $ &  $.074 \pm 0.0012 $ \\  
    \midrule
    \multirow{4}{*}{JointGAN}
    &\multirow{1}{*}{R100k + S50k  }  &  $.798 \pm 0.0012 $   & $.814 \pm 0.0023 $  &  $.719 \pm 0.0023 $    &  $.259 \pm 0.0019 $    &  $.201 \pm 0.0024 $  & $.098 \pm 0.0025 $  & $.079 \pm 0.0031 $  \\
    &\multirow{1}{*}{R100k + S100k }  &  $.812 \pm 0.0013 $   & $.835 \pm 0.0021 $  &  $.725 \pm 0.0019 $    &  $.261 \pm 0.0014 $    &  $.207 \pm 0.0029 $  & $.114 \pm 0.0022 $  & $.081 \pm 0.0034 $ \\
    &\multirow{1}{*}{R100k + S300k }  &  $.826 \pm 0.0018 $  &  $.839 \pm 0.0011 $  &  $.748 \pm 0.0017 $    &  $.272 \pm 0.0015 $    &  $.213 \pm 0.0021 $  & $.129 \pm 0.0029 $  & $.085 \pm 0.0051 $ \\
    &\multirow{1}{*}{R100k + S600k }  &  $.831 \pm 0.0009 $   & $.846 \pm 0.0019 $  &  $.773 \pm 0.0022 $    &  $.313 \pm 0.0011 $    &  $.224 \pm 0.0041 $  & $.134 \pm 0.0021 $  & $.093 \pm 0.0021 $  \\
    \midrule
     \multirow{4}{*}{CoGAN}
    &\multirow{1}{*}{R100k + S50k  }  & $.827  \pm 0.0034 $ & $.843 \pm 0.0019 $    &   $.703 \pm 0.0019  $     &  $.231 \pm 0.0022 $ &  $.192 \pm 0.0012 $ & $.082  \pm 0.0019 $ & $.073 \pm 0.0031 $  \\
    &\multirow{1}{*}{R100k + S100k }  & $.829 \pm 0.0011 $ & $.854  \pm 0.0021 $    &   $.692 \pm 0.0014 $      &  $.214 \pm 0.0023 $ &  $.187 \pm 0.0012 $ & $.073 \pm 0.0022 $  & $.067 \pm 0.0033 $ \\
    &\multirow{1}{*}{R100k + S300k }  & $.831 \pm 0.0013 $ & $.857  \pm 0.0024 $    &   $.724 \pm 0.0015  $     &  $.241 \pm 0.0024 $ &  $.211 \pm 0.0012 $ & $.091 \pm 0.0041 $  & $.076\pm 0.0025 $ \\
    &\multirow{1}{*}{R100k + S600k }  & $.837 \pm 0.0011 $ & $.849  \pm 0.0023 $    &   $.734 \pm 0.0022 $      &  $.251 \pm 0.0021 $ &  $.236 \pm 0.0012 $ & $.114 \pm 0.0032 $  & $0.081 \pm 0.0019 $  \\
    \midrule
    \multirow{4}{*}{SMGAN}
    &\multirow{1}{*}{R100k + S50k  }  &  $.818  \pm 0.0013 $   &  $.832  \pm 0.0013 $    &  $.713  \pm 0.0031 $      &  $.251  \pm 0.0019 $    &   $.203  \pm 0.0034 $     & $.093  \pm 0.0021 $  & $.077 \pm 0.0014 $  \\
    &\multirow{1}{*}{R100k + S100k }  &  $.823  \pm 0.0035 $   &  $.831  \pm 0.0014 $    &  $.723   \pm 0.0032  $    &  $.258  \pm 0.0031 $    &   $.207  \pm 0.0034 $     &  $.098 \pm 0.0022 $   & $.079  \pm 0.0019 $ \\
    &\multirow{1}{*}{R100k + S300k }  &  $.821  \pm 0.0021 $   &  $.847   \pm 0.0019 $   & $.731    \pm 0.0033 $     &  $.263  \pm 0.0018 $    &   $.212  \pm 0.0034 $     &  $.106  \pm 0.0032 $  & $.089  \pm 0.0036 $\\
    &\multirow{1}{*}{R100k + S600k }  &  $.842  \pm 0.0029 $   &  $.859   \pm 0.0018 $   &  $.752   \pm 0.0039 $     &  $.275  \pm 0.0011 $    &   $.236  \pm 0.0034 $     &  $.125  \pm 0.0031 $ & $.096  \pm 0.0025 $  \\
    \midrule
     \multirow{4}{*}{\mname}
    &\multirow{1}{*}{R100k + S50k  }  & $ .835  \pm 0.0015 $    & $.857  \pm 0.0024 $  & $.731  \pm 0.0031 $    &  $.276 \pm 0.0034 $   &  $.204  \pm 0.0027 $  & $.112  \pm 0.0032 $ &  $.078  \pm 0.0021 $ \\
    &\multirow{1}{*}{R100k + S100k }  & $ .842  \pm 0.0021 $    & $.864  \pm 0.0024 $  & $.752  \pm 0.0024 $    &  $.297 \pm 0.0041 $   &  $.216  \pm 0.005 $   & $.132  \pm 0.0019 $ &  $.083  \pm 0.0024 $ \\
    &\multirow{1}{*}{R100k + S300k }  & $ .853   \pm 0.0019 $   & $.869  \pm 0.0028 $  & $.763  \pm 0.0035 $    &  $.324 \pm 0.0042 $   &  $.229  \pm 0.0014 $  & $.145  \pm 0.0014 $ &  $.097  \pm 0.0031 $ \\
    &\multirow{1}{*}{R100k + S600k }  & $ .873   \pm 0.0025 $   & $.884  \pm 0.0026 $  & $.783  \pm 0.0043 $    &  $.346 \pm 0.0022 $  &   $.247  \pm 0.0019 $  & $.169  \pm 0.0018 $ &  $.132  \pm 0.0052 $  \\

    \bottomrule
  \end{tabular}}
  \vspace{-.20cm}
\end{table}


\subsection{Joint Image and Text Evaluation: Can \mname generate high quality pairs of image and reports? }

One of the primary advantages of \mname is the ability to jointly generate paired X-ray images and reports. 
We performed two different experiments to understand the effectiveness of \mname towards generating paired images and reports.

\textbf{Report Generation Task}: X-ray report generation is one of the key tasks in radiology clinical workflow~\cite{schwartz2011improving}.  We validate the effectiveness of augmented paired image and report datasets for report generation task. In this setup, we train report generation models on real data and a combination of real and synthetic data. These trained models are evaluated on held-out real paired datasets. We present the results of these experiments in Table~\ref{tab:data_augmentation_exp}. In this setup, we vary the amount of synthetic data added to the real dataset. We present the performance of real and augmented datasets for report generation task in terms of natural language processing metrics such as CIDEr, BLEU 1-4 \cite{vedantam2015cider,papineni2002bleu}. We show that \mname improves up to $6.9\%$ compared to models trained only on real datasets. This highlights the fact that \mname can be used to augment and improve report generation models.

\textbf{Multimodal joint embeddings of X-ray images and reports}: The multimodal embeddings learned can be used for classification tasks. We perform an experiment to evaluate the joint quality of images and generated text. In table \ref{tab:comparison_multi_emb}a, we compare the result of varying combinations of real and synthetic data on the joint modeling task. In this joint modeling task, we combine features from X-ray images and text reports together for downstream classification. We classify different disease phenotypes using these joint embeddings. We find that adding a synthetic dataset to the real dataset for this joint embedding significantly improves the performance of the classification model.  

\begin{table}[]
    \label{tab:comparison_multi_emb}
   
        \caption{Comparative evaluation of phenotype classification via joint embedding with real and augmented data}
    
          \begin{tabular}{clcc}
            \toprule
            Method & Dataset &   AUC & Acc \\
            \midrule
            \multirow{1}{*}{Only Real }
            &\multirow{1}{*}{Real [100k ]} &   $.849 \pm 0.0025 $ & $.868 \pm 0.0021$  \\  
            \multirow{1}{*}{JointGAN}
            &\multirow{1}{*}{R100k + S300k }   & $.869 \pm 0.0023$  & $.905 \pm 0.0023$   \\
            \multirow{1}{*}{CoGAN}
            &\multirow{1}{*}{R100k + S300k }  & $.871  \pm 0.0014$ & $.896  \pm 0.0019$  \\
            \multirow{1}{*}{SMGAN}
            &\multirow{1}{*}{R100k + S300k }  &  $.883 \pm 0.0016$  & $.902  \pm 0.0015$ \\
            \multirow{1}{*}{\mname}
            &\multirow{1}{*}{R100k + S300k }  & \bf{$.905$} $\pm 0.0019$   & \bf{.924}  $\pm 0.0012$  \\
            \bottomrule
          \end{tabular}
   
\end{table}

\begin{table}[]
    \centering
    \caption{Comparison of generative models with limited labels}
    \begin{tabular}{clccc}
                \toprule
                Method &  Acc & BLEU-1 & FID  \\
                \midrule
                {CoGAN(full)}       &  $0.827 \pm 0.0011$  & $0.247 \pm 0.0023$ &  $15.23 \pm 0.0015$  \\ 
                {JointGAN(full)}    &  $0.813 \pm 0.0014$ & $0.221 \pm 0.0031$ & $16.58   \pm 0.0019$ \\ 
                {SM-GAN(full)}      &  $0.813  \pm 0.0021$ & $0.221 \pm 0.0028$ & $16.58  \pm 0.0021$ \\ 
                {\mname(30\%)}      &  $0.838 \pm 0.0013$ & $0.258 \pm 0.0021$ & $12.84 \pm 0.0021$ \\ 
                {\mname(50\%)}      &  $0.842 \pm 0.0008$ & $0.269 \pm 0.0024$ & $11.73 \pm 0.0023$ \\ 
                {\mname(100\%)}     &  $0.845 \pm 0.0024$ & $0.271 \pm 0.0014$ & $11.31 \pm 0.0028$  \\ 
                \bottomrule
              \end{tabular}
    \caption{Caption}
    \label{tab:my_label}
\end{table}

\subsubsection{Limited label setup: Can we learn a high quality generative model from limited data?}

Machine Learning applications in clinical domains are often limited by the amount of available data and labels. Since generative models require large amounts of data and labels to train, it is a challenge in clinical tasks to learn a high-quality generative models. We show in the following experiments that we can employ self-supervision to overcome the label limitations. We explore the limits of usage of labels by varying the percentage of labels used in the models. In this experiment, we use limited labels ranging from $30\%$, $50\%$ to compare with $100\%$ label usage. We show that even with limited labels \mname can perform competitively. We compare existing baselines to our model which uses self-supervision to able to generate images from limited labels. Table \ref{tab:comparison_multi_emb}b shows that \mname outperforms the baselines in terms of image generation diversity as measured by FID. 

\subsubsection{Case Study: COVID-19 X-ray data augmentation experiment}
We applied the generative models towards improving COVID X-ray classification. In this task, we use \mname to augment chest X-ray images to improve COVID-19 detection. Currently, COVID X-rays classification includes four classes: normal, bacterial pneumonia, viral pneumonia and COVID-19. In this experiment, we evaluate if \mname generated synthetic data can augment chest X-ray image samples for the COVID-19 classification task. Specifically, we compare three different models: \textbf{COVID-19 dataset}, \textbf{pretrained model on CheXpert dataset}, \textbf{pretrained model on combined data of real and synthetic data} ~\cite{cohen2020covid, irvin2019chexpert}. We trained three different models on these datasets. Models pretrained on real dataset and combined dataset are finetuned on the COVID-19 dataset. We show that augmenting real datasets with \mname generated samples improves the overall performance in Table~\ref{tab:phenotype_classification_covid}.

\begin{table}[h!]
\centering
  \caption{Comparison of performance for COVID-19 classification}
  \label{tab:phenotype_classification_covid}
  \resizebox{0.8\columnwidth}{!}{
  \begin{tabular}{c|lccc}
    \toprule
    Type & Phenotype & AUC & Sensitivity  & PPV \\
    \midrule
    \multirow{4}{*}{COVID samples}
    &\multirow{1}{*}{Normal Lung}            &   $0.853 \pm 0.0023$ &  $0.705 \pm 0.0034$ & $0.834 \pm 0.0021$ \\ 
    &\multirow{1}{*}{Bact. Pneumonia }       &   $0.847  \pm 0.0012$ & $0.734 \pm 0.0014$ & $0.768 \pm 0.0041$ \\
    &\multirow{1}{*}{Viral Pneumonia}        &   $0.841  \pm 0.0034$ & $0.758 \pm 0.0015$ & $0.725  \pm 0.0021$ \\   
    &\multirow{1}{*}{COVID-19}               &   $0.853  \pm 0.0033$ & $0.765 \pm 0.0019$ & $0.854 \pm 0.0024$ \\   
    \midrule
    \multirow{4}{*}{ChexPert real dataset}
    &\multirow{1}{*}{Normal Lung}            &   $0.932  \pm 0.0027$ & $0.761  \pm 0.0031$ & $0.931  \pm 0.0014$ \\ 
    &\multirow{1}{*}{Bact. Pneumonia }       &   $0.923 \pm 0.0021 $ & $0.823  \pm 0.0017$ & $0.834   \pm 0.0019$ \\
    &\multirow{1}{*}{Viral Pneumonia}        &   $0.917  \pm 0.0018$ & $0.832  \pm 0.0014$ & $0.745  \pm 0.0009$ \\   
    &\multirow{1}{*}{COVID-19}               &   $0.921  \pm 0.0014$ & $0.863  \pm 0.0021$ & $0.928  \pm 0.0016$ \\   
    \midrule
    \multirow{4}{*}{ChexPert real data + \mname(250k Samples)}
    &\multirow{1}{*}{Normal Lung}            &   $\bf{0.956 } \pm 0.0019$     &  $\bf{0.783 } \pm 0.0017$ & $\bf{0.963  }  \pm 0.0015 $\\ 
    &\multirow{1}{*}{Bact. Pneumonia }       &   $\bf{0.945}   \pm 0.0033$     &  $\bf{0.851}  \pm 0.0014$ & $\bf{0.842  } \pm 0.0018 $ \\
    &\multirow{1}{*}{Viral Pneumonia}        &   $\bf{0.948 }  \pm 0.0031$    &  $\bf{0.853 }  \pm 0.0023$ & $\bf{0.775  } \pm 0.0019 $\\   
    &\multirow{1}{*}{COVID-19}               &   $\bf{0.953 }  \pm 0.0038$  &  $\bf{0.898  }   \pm 0.0034$ & $ \bf{0.943  } \pm 0.0021 $ \\  
    \bottomrule
  \end{tabular}}
  \vspace{-0.20cm}
\end{table}

\subsubsection{Evaluation by Radiologists}
\begin{figure*}[!htb]
\centering
    \includegraphics[width=0.8\textwidth]{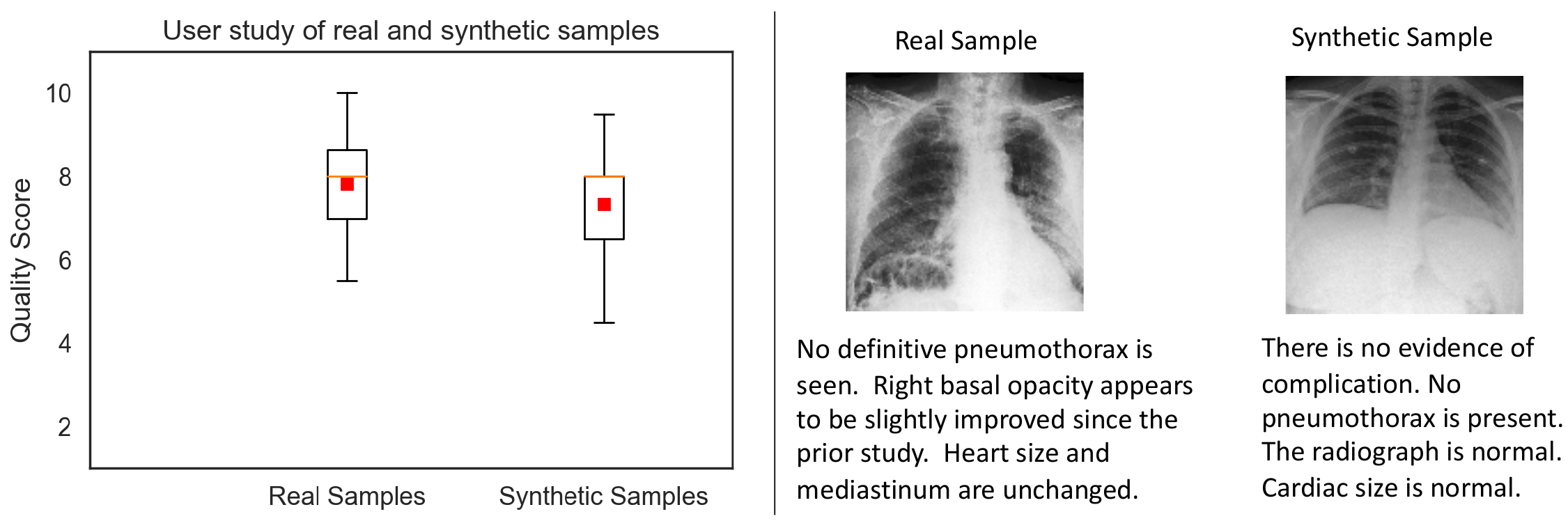}
    \caption{Qualitative evaluation. (a) User study Results (b) Comparative real and synthetic samples}
    \label{fig:qualitative_analysis}
\end{figure*}

We perform a qualitative evaluation of the generated X-ray images and reports. In this task, we present randomized X-ray images and reports to expert doctors. Two radiologists provide a rating between 1-10 for each pair of images and reports. We have presented the results of this evaluation task in figure \ref{fig:qualitative_analysis}. The scores for real and synthetic X-rays samples were  $7.825\pm1.17$ and $7.34\pm1.321$. The inter-rater agreement was $0.832$ measured using cohen's kappa. The comments provided by the doctors indicate that synthetic samples were similar to real samples with some language incoherence in X-ray reports. 

\section{Conclusion}

In this paper, we address the challenging multimodal paired x-ray image and report generation task by proposing a novel self-supervised multimodal generative model called \mname. \mname successfully uses a multimodal generative model to learn to generate paired x-ray images and reports. We use self-supervision to guide \mname to learn from limited samples which are very applicable in the medical domain as the number of labels is often limited. We also use multiple discriminators to guide the process of image generation, report decoding. We show via extensive experiments that \mname can augment real x-ray image datasets to improve downstream classification tasks. Finally, in timely case-study, we show that \mname can also improve COVID-19 x-ray classification.



\medskip

\newpage
\newpage
 
\section*{Broader Impact}
Our paper presents an end-to-end multimodal X-ray generation algorithm to produce synthetic but realistic X-ray images and the corresponding text reports. 

\textbf{Application and societal impact:} 
Deep learning models have shown great promises in medical imaging applications such as automatic diagnosis of radiology images. However, large amount of labeled training data are required to develop  accurate models. Unfortunately, medical data are extremely difficult to share due to the sensitive nature of the data around patient privacy and legal constraints. In addition, many conditions and situations are intrinsically rare which mean limited data. Our proposed method \mname can alleviate these challenges by producing realistic but synthetic data to support model building and augmenting to the limited existing data in some situations as we demonstrated in the COVID-19 image classification task.

As sensing technology become cheap and ubiquitous (e.g., high-resolution cameras from smart phones), it is foreseeable that AI supported telemedicine can efficiently support many people especially the ones in the rural community, where our proposed algorithm can play an important role. 

\textbf{Caveat and potential weakness:}
Although synthetic data can potentially alleviate the sensitive data sharing in healthcare, it is important to study and quantify the privacy implication of synthetic generated data by a model trained with real data. Although unlikely, some real data can be potentially remembered and resynthesized in the synthetic data. There is a balance between data utility and privacy preservation in this line of research.
Finally, a broader trend to consider is that AI technology has largely enabled automation and improved efficiency of many industries such as traditional retail to e-commerce, automation in production plants. The traditional workforce can be negatively impacted. It is important to consider social impact of AI technology to the existing industries. Although in healthcare the skilled experts are still in shortage, the AI based medical technology will probably have limited negative impact in the existing workforce. 
\small
\bibliographystyle{unsrt}
\bibliography{references}
\newpage
\section{Supplementary}

\subsection{Preliminaries: Generative Adversarial Networks}
The Generative Adversarial Network (GAN) involves a Generator ($\operatorname{G}$) and a Discriminator ($\operatorname{D}$) network. The purpose of Generator ($\operatorname{G}$) is to map random noise to samples, while the Discriminator ($\operatorname{D}$) classifies real and generated samples. The generator builds a mapping function from a prior noise distribution $p_{z}(z)$ to data space as $\operatorname{G}(\mathbf{z})$ to learn a generator distribution $p_{g}$, while the discriminator $\operatorname{D}(\mathbf{x})$ outputs a single scalar representing the probability that $\mathbf{x}$ came form training data rather than $p_{g}$ where $p_{\text {data }}$ is the real data distribution. The basic GAN objective function seeks a Nash equilibrium to the following two player min-max problem where value function is defined as \( \min _{G} \max _{D} V(D, G) = \mathbb{E}_{x \sim p_{\text {data }}(\boldsymbol{x})}[\log D(\boldsymbol{x})]+\mathbb{E}_{\boldsymbol{z} \sim p(\boldsymbol{z})}[\log (1-D(G(\boldsymbol{z})))]\) ~\cite{goodfellow2014generative}  where $z \in \mathbb{R}^{d_{z}}$ is a latent variable drawn from distribution $p(\boldsymbol{z})$ such as the unit Gaussian $\mathcal{N}(0, I)$ or the unit uniform $\mathcal{U}[-1,1]$.  Generative adversarial networks can be extended to conditional versions if the generators and discriminators are conditioned on label information $\mathbf{y}$\cite{mirza2014conditional}. The condition information $\mathbf{y}$ and $p(\boldsymbol{z})$ are combined in the joint representation of the generator. The discriminator is provided with generated samples and labels $\mathbf{y}$ as inputs. The objective function can be modified as \(\min _{G} \max _{D} V(D, G)=\mathbb{E}_{\boldsymbol{x} \sim p_{\text {data }}(\boldsymbol{x})}[\log D(\boldsymbol{x} | \boldsymbol{y})]+\mathbb{E}_{\boldsymbol{z} \sim p_{\boldsymbol{z}}(\boldsymbol{z})}[\log (1-D(G(\boldsymbol{z} | \boldsymbol{y})))] \)

\subsection{\mname: Architecture Details}

\subsubsection{Notations Table}
We used these notations to describe different modules. The notations are described in table 4.

\begin{table}[!h]
\centering
  \label{tab:basic_symbols}
  \begin{tabular}{cc}
    \toprule
    Symbol & Definition and description \\
    \midrule
    $\mathbf{I_{n}}$ & Notation for X-ray Images\\
    $\mathbf{S_{n}}$ & Notation for sentences in the X-ray report\\
    $\mathbf{\hat{I}_{n}}$ & Generated X-ray images\\
    $\mathbf{\hat{S}_{n}}$ & Generated X-ray reports\\
    $\mathcal{E}$ & Dataset consisting of images, reports and labels \\
    $\mathbf{y_{n}}$. & Notation for labels associated with images \\
    $\mathbf{w_{N_{s}}}$ & Words in the sentences of X-ray report \\
    $\mathbf{z}$ & Noise vector for the generator \\
    $\operatorname{G(.)}$ & Generator Neural Network \\
    $\operatorname{D_{\text{image}}(.)}$ & X-ray image discriminator Neural Network \\
    $\operatorname{D_{\text{report}}}(.)$ & Discriminator Neural Network \\
    $\operatorname{D_{\text{joint}}}(.)$ & Discriminator Neural Network \\
    $\operatorname{F(.)}$ & Report Generator Network \\
  \bottomrule
\end{tabular}
  \caption{Notations used in \mname}
\end{table}

\subsubsection{\mname Model}
In this section, we provide further description of the different neural networks within in \mname. 

\begin{figure*}[!htb]
\centering
    \includegraphics[width=0.4\textwidth]{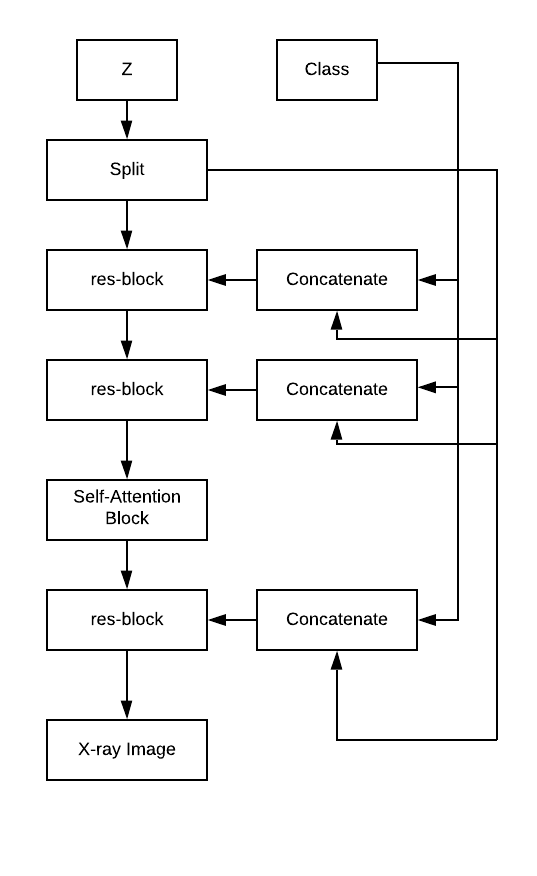}
    \caption{Architectural layout of \mname image generator $\operatorname{G}$}
    \label{fig:generator_network}
\end{figure*}

\textbf{X-ray Image Generator($\operatorname{G}$)}:
Figure \ref{fig:generator_network} shows the architecture of the image generator. X-ray image generator accepts two inputs: (a) noise vector $\boldsymbol{z} \in \mathbb{R}^{120}$ (b) class information $\mathbf{y}$ represented as one-hot vector. 
We embed the class information $\mathbf{y}$ via a linear layer to obtain vector $\mathbf{y_{emb}} \in \mathbb{R}^{128}$. It has been shown generators can use the latent space to influence features at different resolutions by providing direct connections from noise vector to different layers of the generator. We split the noise vector $\mathbf{z}$ to obtain different smaller vectors $\mathbf{z_{spl}} \in \mathbb{R}^{20}$ (https://pytorch.org/docs/master/generated/torch.split.html). The vectors $\mathbf{z_{spl}}$ is passed through a linear layer to obtain $\mathbf{z_{in}}$, $\mathbf{z_{in}} = \operatorname{fc(\mathbf{z_{spl}})}$. We concatenate $\mathbf{z_{in}}$ with $\mathbf{y_{emb}}$ which is passed through three layers of $\operatorname{Res-block-up}$ ~\cite{he2016deep}. We have provided the details of this convolution block in table \ref{tab:res-block-up}. We use $h$, $w$ to denote input height and width and $c_{i}$, $c_{o}$ are input and output channels for the $\operatorname{Res-block-up}$. The output from the previous layer and the concatenated vector from  $\mathbf{z_{in}}$,  $\mathbf{y_{emb}}$ is provided as input to each of the residual block. The final residual output $\mathbf{res_{out}}$ is passed through a self-attention block which applies applies a $1 \times 1$ convolution operation with softmax to obtain intermediate feature vectors which are combined with the original input to compute the $\operatorname{att-block}$, \( \mathbf{\mathbf{att_{out}}}=\operatorname{self-att-block}(\mathbf{res_{out}})\). Finally this output $\operatorname{self-att-block}$ is passed through another $\operatorname{res-block}$ to obtain the $\hat{\mathbf{I}}$ as the output of generator.

\begin{table}[]
\centering
\caption{Details of $\operatorname{Res-block-up}$ for generator}
\begin{tabular}{lll}
\hline Layer & Kernel   & Output \\
\hline Shortcut & {[1,1,1]} &  $2 h \times 2 w \times c_{o}$ \\
\hline $\mathrm{condBN}, \mathrm{ReLU}$ & $-$ &   $h \times w \times c_{i}$ \\
$\mathrm{Conv}$ & {[3,3,1]} &  $2 h \times 2 w \times c_{o}$ \\
$\mathrm{condBN}, \mathrm{ReLU}$ & $-$ &  $2 h \times 2 w \times c_{o}$ \\
$\mathrm{Conv}$ & {[3,3,1]} & $2 h \times 2 w \times c_{o}$ \\
\hline Addition & $-$ &  $2 h \times 2 w \times c_{o}$ \\
\hline
\end{tabular}
\label{tab:res-block-up}
\end{table}

\begin{figure*}[!htb]
\centering
    \includegraphics[width=1\textwidth]{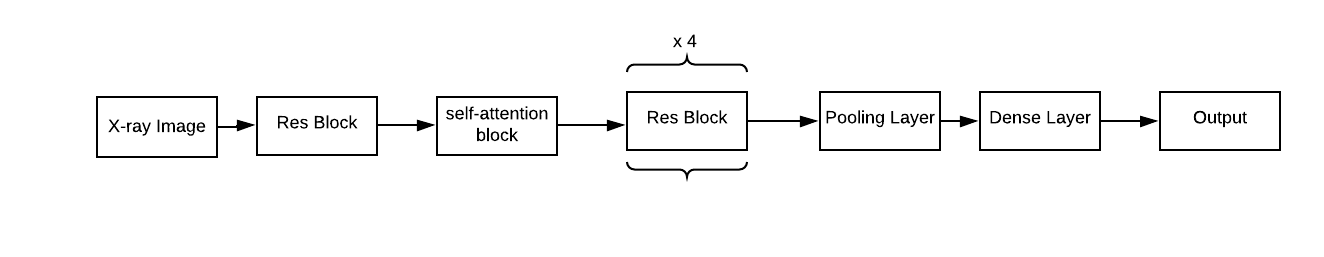}
    \caption{Architectural layout of \mname image generator $\operatorname{D_{\text{Image}}}$}
    \label{fig:discriminator_network}
\end{figure*}

\textbf{X-ray Image Discriminator ($\operatorname{D_{image}}$)}:
Figure \ref{fig:discriminator_network} shows the architecture of the X-ray image discriminator. X-ray image discriminator is used to distinguish between real and fake X-ray images. The discriminator takes an X-ray image $\mathbf{I} \in \mathbb{R}^{ 128 \times 128 \times 3}$ as an input. Image $\mathbf{I}$ is passed through multiple layers of residual convolutional blocks $\operatorname{Res-block-Down}$. We have provided the details of the convolution block in table \ref{tab:tab:res-block-disc}. We use $h$, $w$ to denote input height and width and $c_{i}$, $c_{o}$ are input and output channels for the $\operatorname{Res-block-Down}$. In each residual convolutional block the number of channels is doubled to process the previous layers input. The intermediate feature vector obtained from the residual blocks is passed through a pooling layer and ReLU activation layer. Finally we combine it with the projected condition vector and pass there through a linear layer to obtain the final output.

\begin{table}[]
\centering
 \caption{Details of $\operatorname{Res-block-Down}$ for discriminator. }
    \begin{tabular}{lll}
    \hline Layer & kernel & Output \\
    \hline Shortcut & {[1,1,1]} & $h / 2 \times w / 2 \times c_{o}$ \\
    \hline ReLU  & $-$ & $h \times w \times c_{i}$ \\
    Conv & {[3,3,1]}  & $h \times w \times c_{o}$ \\
    ReLU & $-$ & $h \times w \times c_{o}$ \\
    Conv & {[3,3,1]} & $h / 2 \times w / 2 \times c_{o}$ \\
    \hline Addition & $-$ &  $h / 2 \times w / 2 \times c_{o}$ \\
    \hline
    \end{tabular}
    \label{tab:res-block-disc}
\end{table}

\begin{figure*}[!h]
\centering
    \includegraphics[width=1\textwidth]{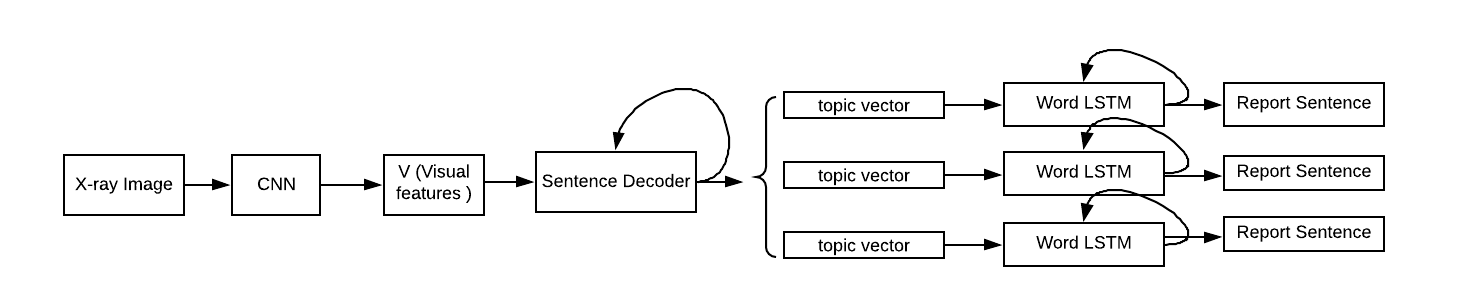}
    \caption{Architectural layout of \mname report generator $\operatorname{F}$}
    \label{fig:report_generation_network}
\end{figure*}

\textbf{X-ray Report Generator ($\operatorname{F}$)}:
We describe the architecture of the X-ray report generation module in Figure \ref{fig:report_generation_network}. The report generation component contains three different sub-components: (a) Image encoder $\operatorname{CNN}$ (b) Sentence $\operatorname{LSTM}$ (c) Word $\operatorname{LSTM}$. The image encoder CNN takes an X-ray image as input and produces feature vectors. This CNN model is pre-trained on X-ray images $\mathbf{I}$ using a DenseNet model. The sentence $\operatorname{LSTM}$ produces topic vectors $\mathbf{t_i}$ which are used as input for word $\operatorname{LSTM}$s to produce the words. After the word $\operatorname{LSTM}$ produces all the words, the words are combined to create the final report $\mathbf{S}$.

\begin{figure*}[!htb]
\centering
    \includegraphics[width=1\textwidth]{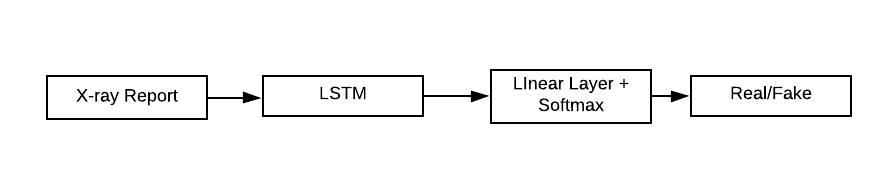}
    \caption{Architectural layout of \mname text discriminator  $\operatorname{D_{\text{Report}}}$}
    \label{fig:text_discriminator}
\end{figure*}

\textbf{X-ray Report Discriminator  ($\operatorname{D_{report}}$)}:
As we show in the figure \ref{fig:text_discriminator}, the X-ray report $\mathbf{S}$ is passed as input to the $\operatorname{LSTM}$. $\operatorname{LSTM}$s have been used to represent paragraphs and sentences to produce context vectors. We use the final representation obtained from the LSTM and pass that to a linear layer. This is finally passed through a softmax layer to obtain the probability of real or fake. 

\begin{figure*}[!htb]
\centering
    \includegraphics[width=1\textwidth]{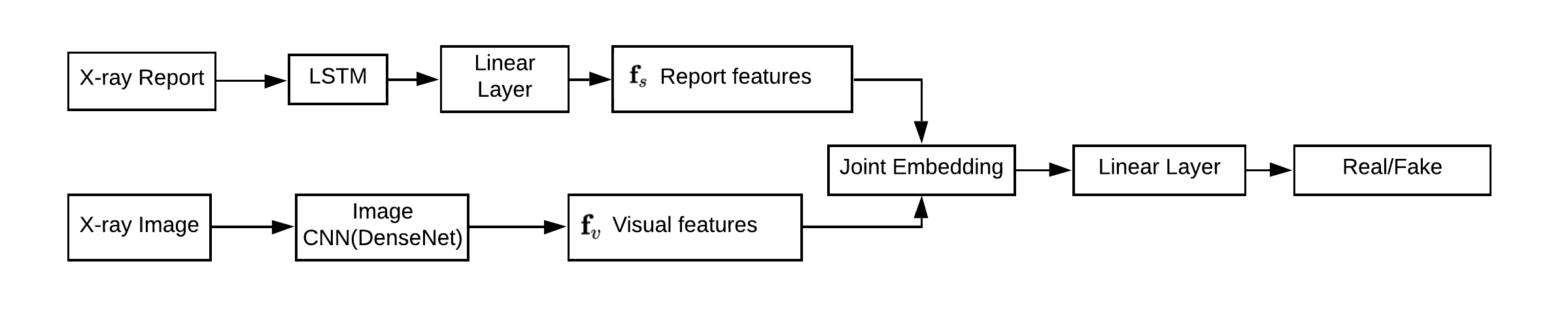}
    \caption{Architectural layout of \mname image generator $\operatorname{D_{\text{Joint}}}$}
    \label{fig:joint_discriminator}
\end{figure*}

\textbf{Joint Discriminator  ($\operatorname{D_{joint}}$)}
As shown in figure \ref{fig:joint_discriminator}, the X-ray report $\mathbf{S}$ and image $\mathbf{I}$ are used to create a joint embedding. X-ray images $\mathbf{I}$ is passed through CNN to obtain an X-ray image feature vector $\mathbf{f_v}$. X-ray report $\mathbf{S}$ is passed through a $\operatorname{LSTM}$ to obtain the final representation of the report $\mathbf{f_s}$. The feature vectors are concatenated together to obtain a joint embedding $\mathbf{C_{\text{joint}}}$. This is finally passed through a linear layer and softmax layer to obtain the probability of embedding being real or fake. 

\subsection{Appendix B Experimental Details}

\subsubsection{Dataset Details}
We used MIMIC-CXR dataset consisting of X-ray images and reports \cite{johnson2019mimic}. This data set was collected from Beth Israel Deaconess Hospital. We apply pre-processing to remove duplicated samples from this dataset. The Radiology reports typically contain an impression and findings section. We extracted the finding section from the report for training our models. We apply tokenization and only keep tokens with at least 6 occurrences in the corpus for training purposes.

\subsubsection{Architecture and hyperparameters}
We use Adam optimizer with a learning rate of $5 \cdot 10^{-5}$ for the generative model and $2\cdot10^{4}$ for the discriminators for training \mname \cite{kingma2014adam}. We staggered discriminator steps and generator steps in 2:1 ratio which led to 400k (800k) generator (discriminator) steps. This helps the discriminator improve it's parameter update process faster compared to a generator. We fix our batch size at 512 while training. We use a noise vector of 120 dimensions as input for the generator. We also use spectral normalization for the layers in the generator and discriminator in the training process. All the models generate $128 \times 128 \times 3 $ X-ray images. We obtain partially labeled data sets for the self-supervised experiments by randomly selecting 30\% of the samples from each class. We rotate the images and use the rotation angles as labels for self-supervision ~\cite{gidaris2018unsupervised}.

\subsubsection{Evaluation Metrics}
\textbf{Fréchet Inception Distance (FID score)}: We first pass real data and generated samples embedded in a specific layer of special pre-trained Inception network on chest X-ray images instead of ImageNet~\cite{heusel2017gans}. Then, a multivariate Gaussian is fit to the data and the distance computed as  $\operatorname{FID}(x,g)=\left\|\mu_{x}-\mu_{g}\right\|_{2}^{2}+\operatorname{Tr}\left(\Sigma_{x}+\Sigma_{g}-2\left(\Sigma_{x} \Sigma_{g}\right)^{\frac{1}{2}}\right)$ where $\mu$ and  $\Sigma$ denote the empirical mean, covariance and subscripts $x$ and $g$ denote the real and generated data respectively.

\subsection{Results}

\subsubsection{Phenotype Classification from X-ray Images with augmented data}
We report the performance of \mname and the baseline models for different phenotype detection from chest X-ray images. The setup for this experiment is similar where we train two models on real X-ray images and generated X-ray images. These trained models are evaluated on held-out X-ray images. The performance of the test X-ray images are reported in Table~\ref{tab:xray_img_clf_classwise}

\begin{table}[h!]
\centering
 \caption{Performance of X-ray image classification using synthetic X-ray}
  \label{tab:xray_img_clf_classwise}
  \resizebox{1\columnwidth}{!}{
  \begin{tabular}{c|lccccc}
    \toprule
    Dataset & Method
    &  Cardiomegaly & Consolidation & Pleural Effusion & Pneumothorax & Pulmonary Edema \\
    \midrule
    \multirow{6}{*}{MIMIC}
    &\multirow{1}{*}{Real data [100k images]}    &  0.812 & 0.847 & 0.753 & 0.735 &  0.732 \\ 
    &\multirow{1}{*}{CoGAN [100k images]}        &  0.741 & 0.817 & 0.708 & 0.713 &  0.682 \\ 
    &\multirow{1}{*}{JointGAN [100k images]}     &  0.732 & 0.785 & 0.724 & 0.681 &  0.713 \\ 
    &\multirow{1}{*}{\mname [100k images]}       &  0.784 & 0.734 & 0.728 & 0.715 &  0.718 \\ 
    \bottomrule
  \end{tabular}}
  \vspace{-0.20cm}
\end{table}

\subsubsection{Performance comparison of augmented data to real data}

We performed an experiment to evaluate augmented datasets in comparison to real datasets of similar size. In this setup, we keep the total size of the dataset constant at 100k and change the ratio of real and synthetic images. We present the results of this experiment in Table \ref{tab:classification_cmp2}. This experiment evaluates the performance of augmented datasets where the total dataset size is low. We show that even when we use $80\%$ fewer real images, augmented datasets only show $6\%$ decrease in performance. This shows that even with low-data availability, synthetic data augmentation can perform competitively compared to models trained only on real X-ray images. 
 
\begin{table}[htb!]
    \centering
    \caption{X-ray image classification performance comparison with \mname augmented data. Dataset size at 100k while reducing the amount of real images in the augmented dataset. In this table, R indicates Real data and S indicate Synthetic data.}
     \begin{tabular}{clcc}
            \toprule
                Method & Data &  AUC & Acc \\
                \midrule
                \multirow{1}{*}{Only Real}
                &\multirow{1}{*}{R100k } &  .824 & .846 \\ 
                \midrule
                \multirow{4}{*}{JointGAN}
                &\multirow{1}{*}{R90k + S10k}  &  .796   & .813 \\
                &\multirow{1}{*}{R80k + S20k}  &  .778  &  .801 \\
                &\multirow{1}{*}{R60k + S40k}  &  .745  &  .764 \\
                &\multirow{1}{*}{R20k + S80k}  &  .717  &  .732 \\
                \midrule
                 \multirow{4}{*}{CoGAN}
                &\multirow{1}{*}{R90k + S10k}  &  .784  & .808 \\
                &\multirow{1}{*}{R80k + S20k}  &  .771  &  .796 \\
                &\multirow{1}{*}{R60k + S40k}  &  .736  &  .757 \\
                &\multirow{1}{*}{R20k + S80k}  &  .712  &  .746 \\
                \midrule
                \multirow{4}{*}{SMGAN}
                &\multirow{1}{*}{R90k + S10k}  & .794  &  .812 \\
                &\multirow{1}{*}{R80k + S20k}  & .764  & .783 \\
                &\multirow{1}{*}{R60k + S40k}  & .742  &  .763 \\
                &\multirow{1}{*}{R20k + S80k}  & .723  &  .742 \\
                \midrule
                \multirow{4}{*}{\mname}
                &\multirow{1}{*}{R90k + S10k}  & .808  & .828 \\
                &\multirow{1}{*}{R80k + S20k}  &  .792  &  .821  \\
                &\multirow{1}{*}{R60k+ S40k}  &  .773  &  .796 \\
                &\multirow{1}{*}{R20k + S80k}  &  .756  &  .774 \\
                \bottomrule
     \end{tabular}
     \label{tab:classification_cmp2}
\end{table}

\subsubsection{Additional generated data samples}

\begin{figure*}[!htb]
\centering
    \includegraphics[width=1\textwidth]{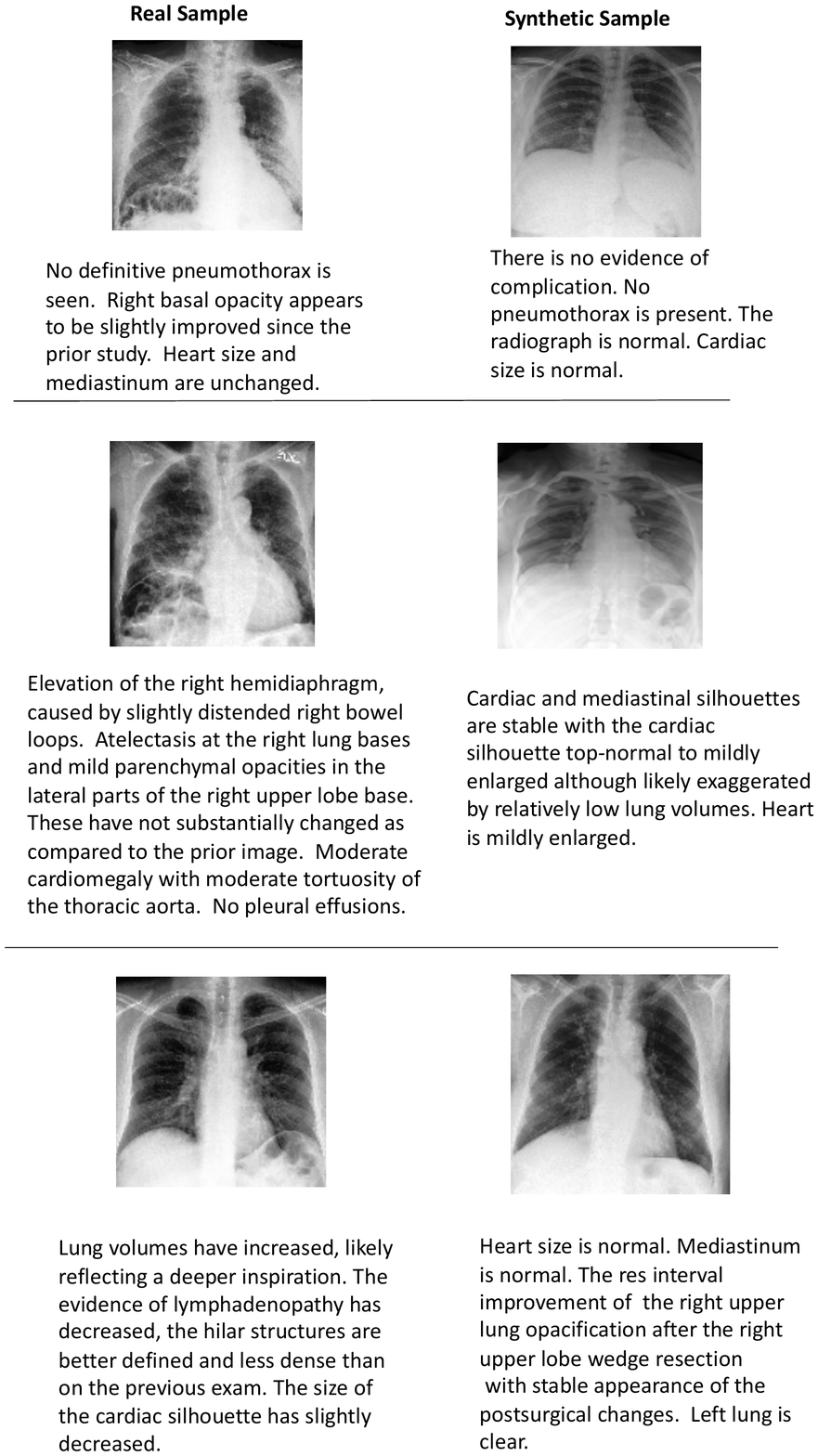}
    \caption{Comparison of Real X-ray image and report pairs with generated X-ray images, reports pairs}
    \label{fig:samples_1_paired}
\end{figure*}

\begin{figure*}[!htb]
\centering
    \includegraphics[width=1\textwidth]{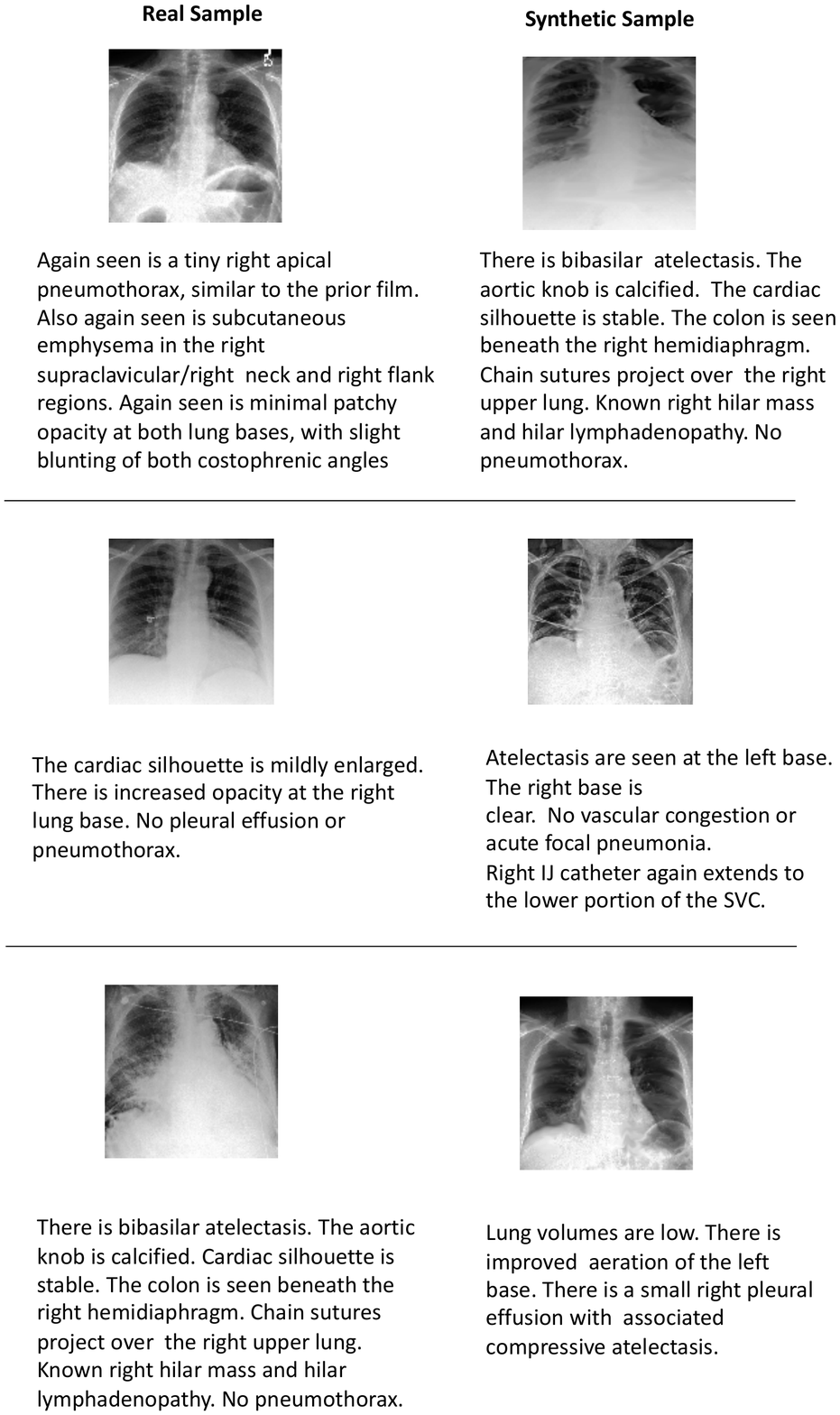}
    \caption{Comparison of Real X-ray image and report pairs with generated X-ray images, reports pairs}
    \label{fig:samples_2_paired}
\end{figure*}

\begin{figure}
\centering
\begin{minipage}{.5\textwidth}
  \centering
  \includegraphics[width=1\linewidth]{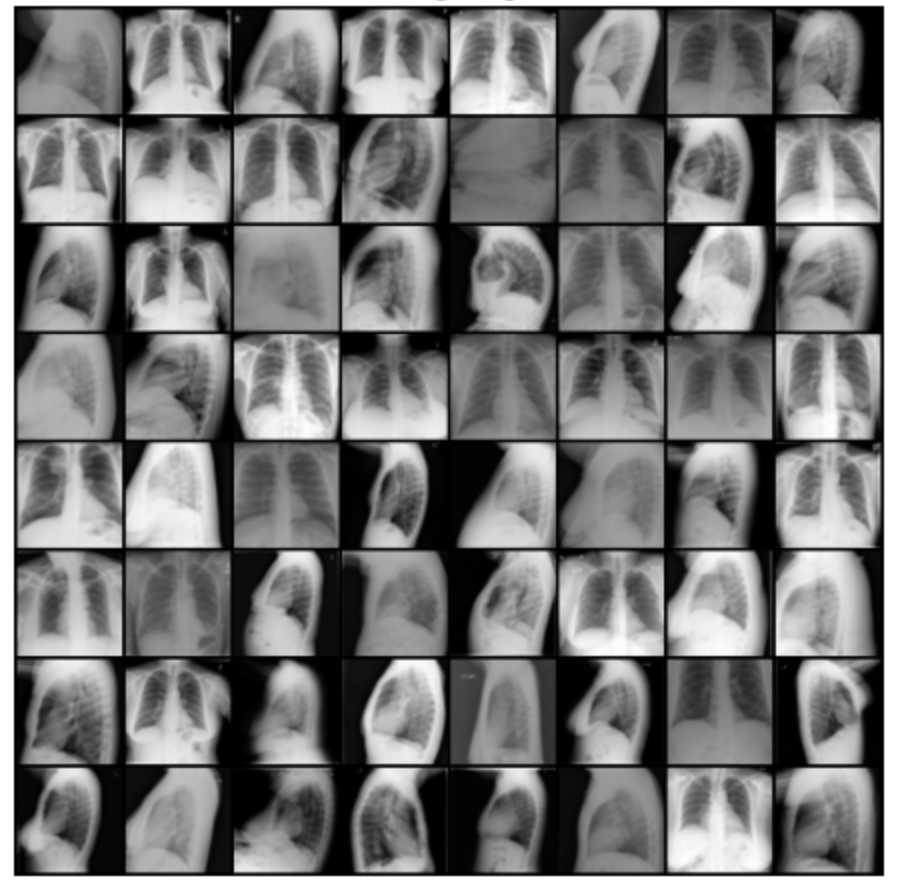}
  \caption{figure}{Real X-ray images}
  \label{fig:real_1}
\end{minipage}%
\begin{minipage}{.5\textwidth}
  \centering
  \includegraphics[width=1\linewidth]{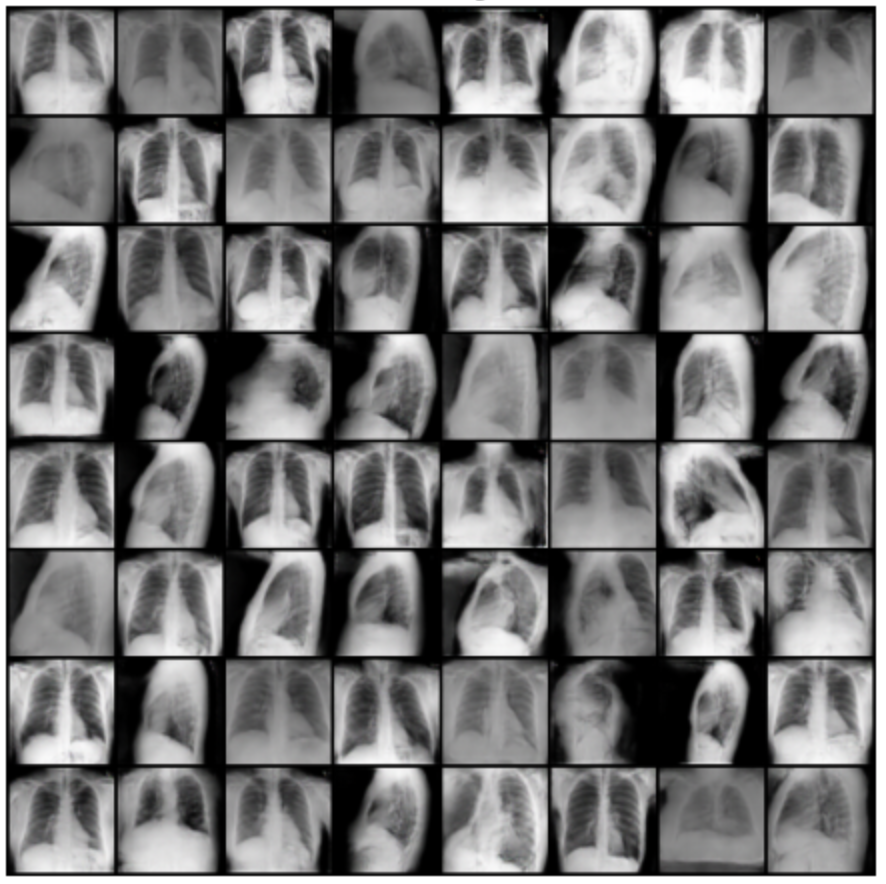}
  \caption{figure}{Synthetic X-ray images}
  \label{fig:samples_4}
\end{minipage}
\end{figure}

\begin{figure*}[!htb]
\centering
    \includegraphics[width=0.8\textwidth]{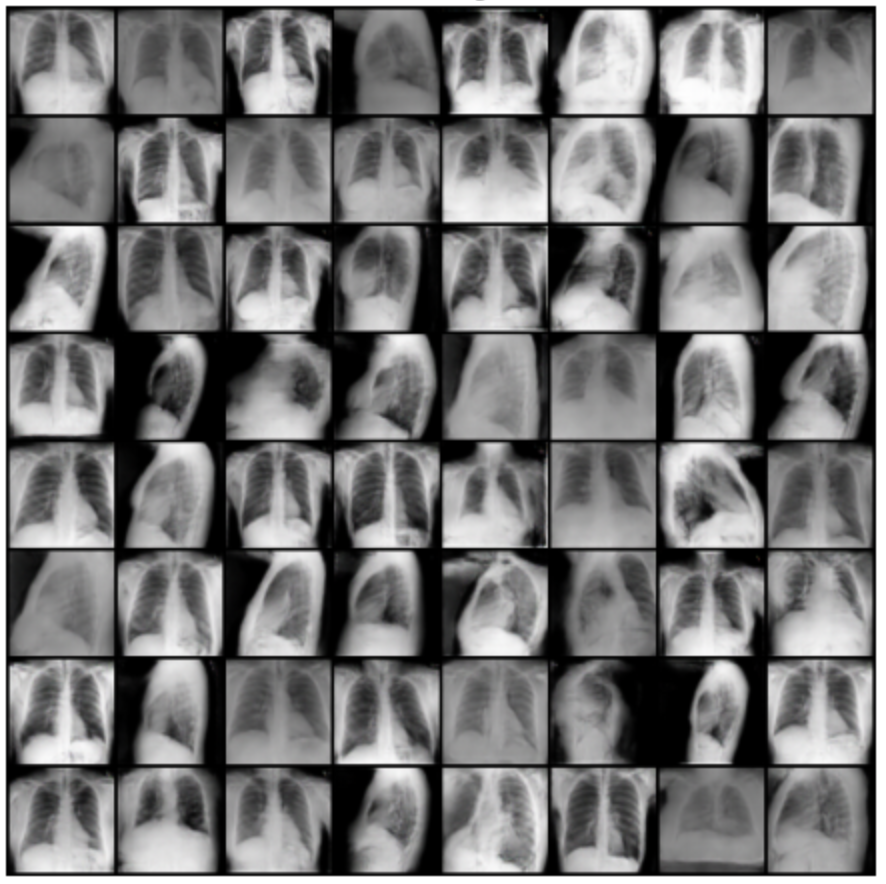}
    \caption{Samples of Synthetic X-ray images}
    \label{fig:samples_5}
\end{figure*}

In figures \ref{fig:samples_1_paired} and \ref{fig:samples_2_paired}, we show additional generated X-ray image,report pairs in comparison to real X-ray image and report pairs. Figure \ref{fig:real_1} and \ref{fig:samples_4} show comparison of real X-ray images to synthetic X-ray images. Finally, figure \ref{fig:samples_5} shows more synthetic X-ray images.

\end{document}